\documentclass[twocolumn,showpacs,prl,superscriptaddress]{revtex4}         
\def\sNN{\mbox{$\sqrt{s_{_{NN}}}$}}  
\def\etal{{\it et al.}} 
\newcommand{ \be }{\begin{equation}}      
\newcommand{ \ee }{\end{equation}}      
\newcommand{ \bea }{\begin{eqnarray}}      
\newcommand{ \eea }{\end{eqnarray}}

\usepackage{color}

\usepackage{graphicx} % Include figure files         
\usepackage{dcolumn} % Align table columns on decimal point         
\usepackage{bm} % bold math         
\begin{document}         
\title{      
\begin{flushright} 
%{\small \sl version 22,  \today \\ 
% }      
\end{flushright} 
Strangelet Search at the BNL Relativistic Heavy Ion Collider 
}

\affiliation{Argonne National Laboratory, Argonne, Illinois 60439}
\affiliation{University of Birmingham, Birmingham, United Kingdom}
\affiliation{Brookhaven National Laboratory, Upton, New York 11973}
\affiliation{California Institute of Technology, Pasadena, California 91125}
\affiliation{University of California, Berkeley, California 94720}
\affiliation{University of California, Davis, California 95616}
\affiliation{University of California, Los Angeles, California 90095}
\affiliation{Carnegie Mellon University, Pittsburgh, Pennsylvania 15213}
\affiliation{University of Illinois, Chicago}
\affiliation{Creighton University, Omaha, Nebraska 68178}
\affiliation{Nuclear Physics Institute AS CR, 250 68 \v{R}e\v{z}/Prague, Czech Republic}
\affiliation{Laboratory for High Energy (JINR), Dubna, Russia}
\affiliation{Particle Physics Laboratory (JINR), Dubna, Russia}
\affiliation{University of Frankfurt, Frankfurt, Germany}
\affiliation{Institute of Physics, Bhubaneswar 751005, India}
\affiliation{Indian Institute of Technology, Mumbai, India}
\affiliation{Indiana University, Bloomington, Indiana 47408}
\affiliation{Institut de Recherches Subatomiques, Strasbourg, France}
\affiliation{University of Jammu, Jammu 180001, India}
\affiliation{Kent State University, Kent, Ohio 44242}
\affiliation{Institute of Modern Physics, Lanzhou, China}
\affiliation{Lawrence Berkeley National Laboratory, Berkeley, California 94720}
\affiliation{Massachusetts Institute of Technology, Cambridge, MA 02139-4307}
\affiliation{Max-Planck-Institut f\"ur Physik, Munich, Germany}
\affiliation{Michigan State University, East Lansing, Michigan 48824}
\affiliation{Moscow Engineering Physics Institute, Moscow Russia}
\affiliation{City College of New York, New York City, New York 10031}
\affiliation{NIKHEF and Utrecht University, Amsterdam, The Netherlands}
\affiliation{Ohio State University, Columbus, Ohio 43210}
\affiliation{Panjab University, Chandigarh 160014, India}
\affiliation{Pennsylvania State University, University Park, Pennsylvania 16802}
\affiliation{Institute of High Energy Physics, Protvino, Russia}
\affiliation{Purdue University, West Lafayette, Indiana 47907}
\affiliation{Pusan National University, Pusan, Republic of Korea}
\affiliation{University of Rajasthan, Jaipur 302004, India}
\affiliation{Rice University, Houston, Texas 77251}
\affiliation{Universidade de Sao Paulo, Sao Paulo, Brazil}
\affiliation{University of Science \& Technology of China, Hefei 230026, China}
\affiliation{Shanghai Institute of Applied Physics, Shanghai 201800, China}
\affiliation{SUBATECH, Nantes, France}
\affiliation{Texas A\&M University, College Station, Texas 77843}
\affiliation{University of Texas, Austin, Texas 78712}
\affiliation{Tsinghua University, Beijing 100084, China}
\affiliation{Valparaiso University, Valparaiso, Indiana 46383}
\affiliation{Variable Energy Cyclotron Centre, Kolkata 700064, India}
\affiliation{Warsaw University of Technology, Warsaw, Poland}
\affiliation{University of Washington, Seattle, Washington 98195}
\affiliation{Wayne State University, Detroit, Michigan 48201}
\affiliation{Institute of Particle Physics, CCNU (HZNU), Wuhan 430079, China}
\affiliation{Yale University, New Haven, Connecticut 06520}
\affiliation{University of Zagreb, Zagreb, HR-10002, Croatia}

\author{B.I.~Abelev}\affiliation{University of Illinois, Chicago}
\author{M.M.~Aggarwal}\affiliation{Panjab University, Chandigarh 160014, India}
\author{Z.~Ahammed}\affiliation{Variable Energy Cyclotron Centre, Kolkata 700064, India}
\author{B.D.~Anderson}\affiliation{Kent State University, Kent, Ohio 44242}
\author{D.~Arkhipkin}\affiliation{Particle Physics Laboratory (JINR), Dubna, Russia}
\author{G.S.~Averichev}\affiliation{Laboratory for High Energy (JINR), Dubna, Russia}
\author{Y.~Bai}\affiliation{NIKHEF and Utrecht University, Amsterdam, The Netherlands}
\author{J.~Balewski}\affiliation{Indiana University, Bloomington, Indiana 47408}
\author{O.~Barannikova}\affiliation{University of Illinois, Chicago}
\author{L.S.~Barnby}\affiliation{University of Birmingham, Birmingham, United Kingdom}
% ...
\collaboration{STAR Collaboration}\noaffiliation

\begin{abstract}  

We have searched for strangelets in a triggered sample of 61 million  
central (top 4\%) Au+Au collisions at $\sNN = 200~$GeV near beam  
rapidities at the STAR detector.  We have sensitivity to metastable  
strangelets with lifetimes of order $\geq 0.1 ns$, in contrast to limits  
over ten times longer in AGS studies and longer still at the SPS.     
Upper limits of a few $10^{-6}$ to $10^{-7}$ per central Au+Au  
collision are set for strangelets with mass ${}^{>}_{\sim}30$  
GeV/$c^{2}$. 
 
\end{abstract}  
\maketitle 
 
Strange Quark Matter (SQM) is a hypothetical state of matter 
consisting of roughly equal numbers of $u$, $d$ and $s$ quarks. It 
might have lower energy per baryon than ordinary nuclear matter and 
thus might be the true ground state of baryonic 
matter~\cite{Bodmer,Witten}. Strangelets might be stable, or 
meta-stable with a weak-decay lifetime.  They are predicted to have a 
small or zero charge-to-mass ratio, and could carry either sign of 
charge.  SQM has been proposed to explain several astrophysical 
phenomena~\cite{Witten,madsen1}, to be used as a clean energy 
source~\cite{energy}, to be a QCD laboratory~\cite{qcd-lab} and to 
cause possible exotic scenarios~\cite{disaster}. 

Terrestrial materials~\cite{Klingenberg}, cosmic remnants~\cite{  
Klingenberg}, and heavy-ion collisions (E864 and E896 at Brookhaven  
and NA52 at CERN)~\cite{PastSearch,liushaw} have been searched for  
strangelets.  All of these searches yielded negative results and  
reported complementary upper limits. 
 
Coalescence~\cite{Baltz} and thermal statistical~\cite{Braun-Munzinger}   
models predict low rates of strangelet production at midrapidity, as 
supported by related measurements of nucleon coalescence~\cite{e864nuclei}.   
If a Quark Gluon Plasma (QGP) is created in heavy-ion collisions, it could  
cool down by distillation (kaon emission) and condense to strange-quark-rich  
matter in its ground state -- a strangelet~\cite{Greiner,Hank}.  However,  
this requires a net baryon excess and a non-explosive process in the 
collisions~\cite{Greiner,sandweiss}, neither of which is favored at  
midrapidity at the Relativistic Heavy Ion Collider (RHIC)~\cite{starwhitepaper}.  

Pomerons -- force carriers invoked in some soft-interaction descriptions  
-- have been introduced as a possible mechanism for strangelet production  
near spectator rapidities; QGP formation is not required~\cite{Bleicher}.    
Models based on color superconductivity predict that quark pairs at 
high baryon density have lower energy when colors and flavors are 
correlated to form color-flavor locked (CFL) pairs~\cite{madsen2}; 
positively-charged strangelets are favored because of the surface 
depletion of $s$ quarks~\cite{madsen2}. $u$ and $d$ outnumber $s$ quarks  
in such states. These scenarios require high baryon density and favor  
forward rapidities, hence the focus of the present search.  We investigate  
the energy and shower profiles in calorimeters at zero degrees relative to  
the beam lines. This approach provides excellent sensitivity for short-lived  
strangelets and has a near-uniform acceptance for strangelets with a wide range of charge-to-mass ratios,  
including neutrals. It complements the majority of past experiments, which
searched for strangelets with the hope that most of the strangelets are 
produced around midrapidity~\cite{PastSearch}.

\begin{figure} 
\begin{center} 
\resizebox{!}{5cm}{ 
\includegraphics{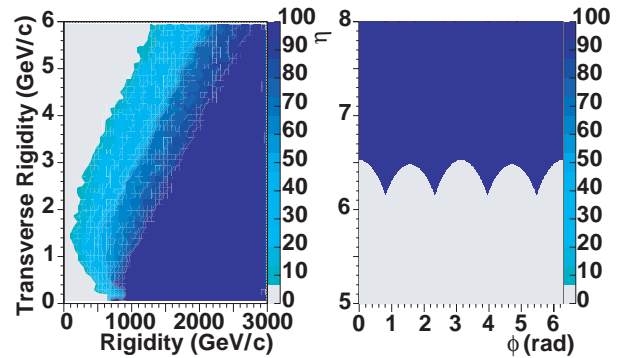}} 
\caption{(Color online) The geometrical acceptance indicated by color  
coding (in \%) for the STAR ZDC-SMD as a function of transverse rigidity  
and rigidity for charged strangelets (left), and as a function of  
pseudorapidity ($\eta$) and azimuthal angle ($\phi$) for neutral  
strangelets (right). 
\label{fig:acceptance}} 
\end{center} 
\end{figure}

A strangelet produced in the forward region at RHIC will have a small  
charge-to-mass ratio and its rigidity (momentum$/Z$, where $Z$ is the  
particle's charge in multiples of the electron charge) will be  
exceptionally large.  Thus it will follow a nearly straight trajectory  
and deposit a large signal in one of the Zero Degree Calorimeters  
(ZDCs)~\cite{ZDC}, which are located just beyond the nearest beam  
dipole magnets (DX) to the experiment.  Normal spectator fragments  
other than neutrons cannot reach the ZDCs due to the strong fields of  
the DX magnets.  There is one ZDC to the east of the interaction region  
at STAR, and another to the west, located $~18$ m from the center of  
the detector.  The digitizing (ADC) gate for the ZDCs is open from  
$-15$ ns to $+70$ ns relative to the arrival time for particles from the  
interaction moving at the speed of light.  The acceptance for a charged  
strangelet in a ZDC depends on its rigidity and, because of the ZDC's  
rectangular shape, depends slightly on its azimuthal angle.  The left  
panel of Fig.\ref{fig:acceptance} shows the acceptance in total rigidity  
and transverse rigidity. The acceptance for a neutral strangelet depends  
on its pseudorapidity $\eta$ ($\equiv - {\ln \tan (\theta/2)}$, where  
$\cos\theta = p_z /p$) and azimuthal angle $\phi$ (Fig.~\ref{fig:acceptance},  
right panel).  
%----------------------------------------------------------------------- 
%----------------------------------------------------------------------- 
 
\begin{figure} 
\begin{center} 
\resizebox{!}{3.6cm}{ 
\includegraphics{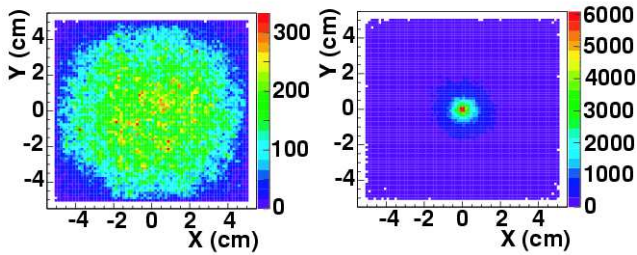}} 
\caption{(Color online) The shower profile of neutron clusters (left) and strangelets 
(right) from simulations. \label{fig:simulation}} 
\end{center} 
\end{figure}

A strangelet would produce a large shower in a ZDC, comparable to 
the signal from a cluster of spectator neutrons.  The latter signal is 
dispersed in the transverse plane, since transverse momentum for 
spectator neutrons is comparable to Fermi momentum.  In contrast, the 
shower from a strangelet would originate at a single point in the ZDC,  
as illustrated by the GEANT simulation in 
Fig.~\ref{fig:simulation}, where the shower profile in the X-Y 
(transverse) plane is plotted for spectator neutrons (left) and 
strangelets (right).  
In this simulation, each neutron 
cluster consists of 35 neutrons with a maximum $p_t$ of 270 MeV/c, 
and each strangelet has the same mass as 35 neutrons, with the assumed 
cross section being the same as that of a neutron times the baryon 
number of the strangelet.  The Root Mean Square (RMS) of the radial 
extent of the strangelet shower is $(69 \pm 12)$\% times the RMS from the 
simulated 35-neutron spectator remnant.  A change in the assumed mass 
of the strangelet or in the number of spectator neutrons does not 
affect this ratio.  For the scenario of a strangelet accompanied by 
neutrons, the mean-square is a linear combination of that from 
strangelet and neutrons. In a case where a strangelet is accompanied 
by 10 neutrons, the ratio increases by 12\%. Therefore the 
strangelet signature used in this ZDC search is a large energy deposition  
with a narrow transverse shower profile in central Au+Au collisions. 
%----------------------------------------------------------------------- 
%----------------------------------------------------------------------- 

Each ZDC incorporates a Shower Maximum Detector (SMD)~\cite{STARSMD}  
with 7 vertical and 8 horizontal slats, providing event-by-event  
information that allows a strangelet signal to be distinguished from  
multiple neutrons via the narrower lateral extent of its shower.   
The strangelet trigger takes advantage of the anti-correlation of 
particle multiplicity and ZDC signal for central heavy-ion 
collisions; in central events, the background ZDC signal from spectator  
neutrons is small and the probability of production of an exotic high-mass object  
is highest.  We take the shower profile of neutrons 
from peripheral events as a reference sample, because of the large ZDC 
signals they produce.  The particle multiplicity used for online 
triggers in STAR is measured primarily with the Central Trigger Barrel 
(CTB) covering $|\eta|<1$ and $0<\phi<2\pi$~\cite{trigNIM}. Although a 
normal central event produces a small signal in the ZDC, a peripheral 
Au+Au collision can accompany the central event in one bunch crossing 
and may result in a background event with large ZDC signal. These 
background double-interaction events happen at a level of $0.1\%$ -- 
$0.01\%$ with current luminosity ($1 \times 10^{27}$ $\rm{cm}^{-2} 
\rm{s}^{-1}$). The ZDC-SMD provides the needed shower profile 
information to distinguish those events from strangelet events.  
 
\begin{figure} 
\resizebox{!}{8.cm}{ 
\includegraphics{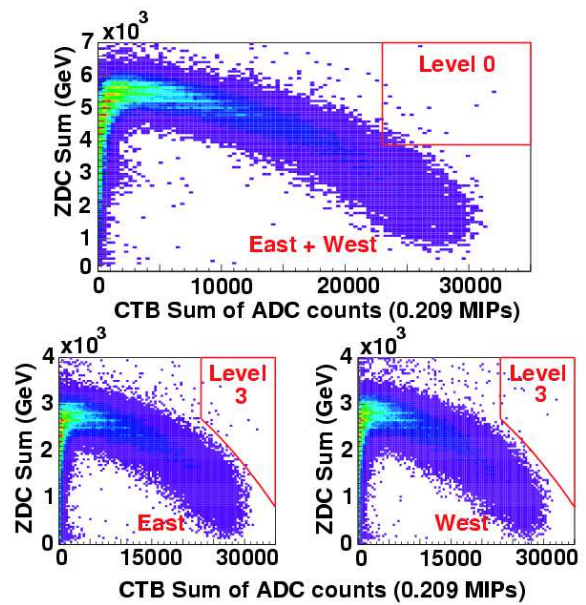}} 
\caption{(Color online) Level 0 and Level 3 triggers. \label{fig:trigger}} 
\end{figure}

During the 2004 run, two special triggers for the strangelet search were 
implemented.  The trigger conditions at level zero (L0) are that the 
signal from the CTB exceeds 23000 ADC counts (approximately 
4800 minimum ionizing particles), which corresponds to selecting the 
top 4\% most central events, and the signal sum from both ZDCs  
exceeds 3875 GeV (39 neutron equivalents);  
see Fig.~\ref{fig:trigger}, upper panel. The total rejection obtained 
with this trigger is 99.25\%.  
%----------------------------------------------------------------------- 
%----------------------------------------------------------------------- 
 
In total, we recorded 458 thousand L0 triggered events, sampling 61 
million 4\% most central Au+Au events during the 2004 run.   
A level-three (L3) trigger was implemented to write 
to an express stream approximately 20\% of the events that passed the 
L0 trigger.  The L3 trigger required that the correlation between  
either one of the ZDCs and the CTB signal must lie above the curve made  
by the correlation observed in minimum-bias events   
(Fig.~\ref{fig:trigger}, 
bottom panels). Our analysis only applies to events that pass the L3 
trigger. Additional offline requirements are  
that the event vertex z-position as 
determined from the ZDCs was within $2\sigma$ of the 
z-position determined from Time Projection Chamber (TPC) tracking, and the energy sum measured 
in the SMDs was within $2\sigma$ of the energy sum 
measured in the ZDCs.  The first requirement removes possible 
events from pile-up, and the second ensures that large signals 
are deposited in the ZDCs and their associated SMDs for triggered 
events.  Finally, since we record the ADC value for each ZDC module,  
as well as the ADC value of the analog sum of the three ZDC modules, 
we can compare these two signals to eliminate any events with an 
electronics-related inconsistency. Events that pass all triggers and 
cuts have a typical energy deposition of 3300 GeV at the ZDC selected 
by the L3 trigger. 
 
\begin{figure} 
  \begin{center} 
\resizebox{!}{4.0cm}{ 
\includegraphics{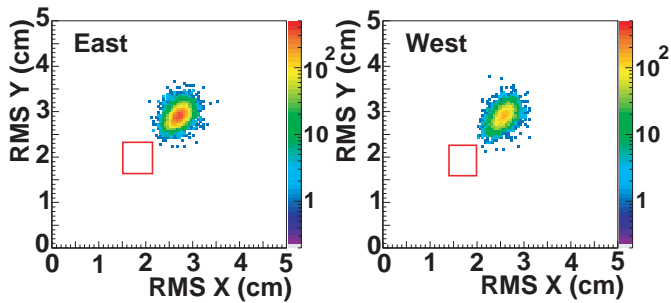}} 
\caption{(Color online) The distribution of RMS from east SMD (left) and west SMD (right).  
\label{fig:rms}} 
  \end{center} 
\end{figure}

We recorded the signal of each SMD slat for each event and computed 
the signal-weighted centroid and variance for each event in both the 
$X$ and $Y$ directions.  Figure~\ref{fig:rms} shows the distribution 
of RMS values from the SMD in the  $X - Y$ plane  
for strangelet candidates surviving the cuts described 
previously. If a strangelet is created in the collision and reaches 
one of the ZDC-SMDs, it is expected to produce a large signal with a 
relatively narrow shower profile in the SMD.  Any candidate with such 
characteristics would show up in the lower-left boxes in one of the 
two panels. The position and the size of the two boxes are obtained 
from the simulation.  
If a different reference sample other than peripheral events is  
used to study the possible mismatch of shower profile, the position and  
width of the search windows change by only about 3\%. 
Figure~\ref{fig:rms} shows that the RMS distributions for both SMDs 
from the strangelet candidates are well above the shower profiles 
expected from strangelets. We conclude that no candidate with 
anomalously low RMS in both $X$ and $Y$ directions is observed. 
%----------------------------------------------------------------------- 
%----------------------------------------------------------------------- 
 
\begin{figure} 
  \begin{center} 
\resizebox{!}{5.5cm}{ 
\includegraphics{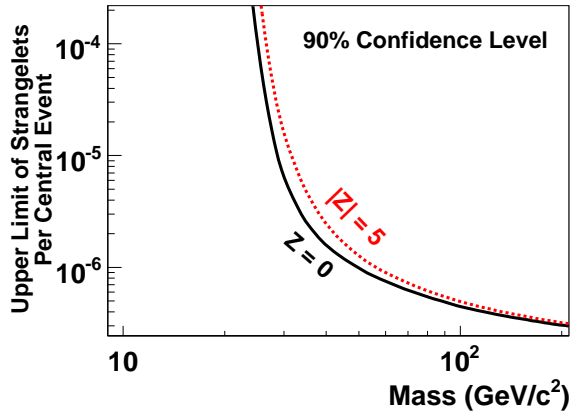}} 
\caption{(Color online) The upper limit, with 90\% confidence level, of strangelet 
production as a function of mass, for neutral strangelets (solid line) 
and strangelets with charge = 5 (dashed line). Limits for  
strangelets with charge 1 -- 4 lie between the two lines.  
\label{fig:upperLimit}} 
  \end{center} 
\end{figure}

To establish an upper limit for strangelet production, we assume a 
source distribution having an inverse slope in transverse mass of 160 MeV,  
which is a typical chemical freezeout temperature at 
RHIC.  Since strangelet production could be enhanced at 
midrapidity due to the formation of a QGP \cite{Greiner}, or enhanced 
at forward rapidity due to the Pomeron-cutting mechanism 
\cite{Bleicher}, we assume a flat distribution in rapidity up to  
beam rapidity. Correcting for our trigger efficiencies and timing gate 
efficiencies, as well as our acceptance, we present our upper limit at 
90\% confidence level as a function of mass in 
Fig.~\ref{fig:upperLimit}. 
%----------------------------------------------------------------------- 
%----------------------------------------------------------------------- 
%----------------------------------------------------------------------- 
\begin{figure} 
  \begin{center} 
\resizebox{!}{5.cm}{ \includegraphics{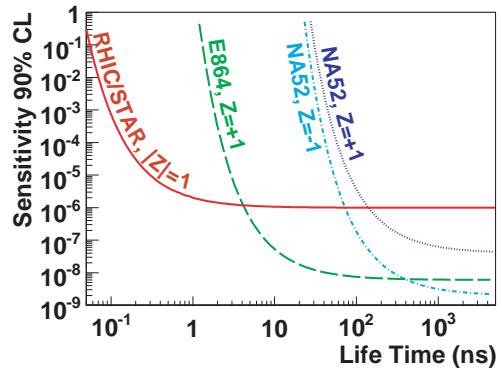}} 
\caption{(Color online) The upper limit of strangelet production at 
90\% confidence level as a function of lifetime for strangelet with 
mass 50 GeV/$c^{2}$, for which all experiments have a sensitivity.  
Solid, dashed, dot-dashed, and dotted 
lines are the limits of $|Z|$ = 1 strangelets from the 4\% most central 
Au+Au collisions at RHIC, Z=+1 strangelets from E864/AGS for the 10\% most 
central Au+Pt collisions, and Z=-1 and Z=+1 strangelets from NA52/SPS 
for minimum-bias Pb+Pb collisions, respectively.  
\label{fig:upperLimitLifeTime}} 
  \end{center} 
\end{figure} 
%----------------------------------------------------------------------- 
 
The upper limits decrease as the assumed strangelet mass increases. 
The limits are $6.9 \times 10^{-6}$ and $4.5 \times 10^{-7}$ per 
central Au+Au collision for neutral strangelets with mass 30 
GeV/$c^{2}$ and 100 GeV/$c^{2}$, respectively.  Limits for charged 
strangelets are different 
because of the 
magnets in front of the ZDC-SMDs. Varying the assumed transverse mass 
slope for strangelets by $\pm$ 60 MeV results in little 
change to the upper limits.  However, the limits depend strongly 
on the rapidity distribution. For the strangelet mass range presented 
here, our acceptance decreases dramatically at 
$y{}^{<}_{\sim}4$ with the detailed geometric acceptance shown in 
Fig.~\ref{fig:acceptance}. This dependence is not uncommon 
because none of the strangelet search experiments have $4\pi$ 
coverage over all the kinematics while the limits are for production 
per interaction in all kinematics. Although our upper limits are roughly 
comparable to those set by other heavy-ion experiments (ranging from 
$10^{-7}$ to $10^{-9}$, as shown in Fig.~\ref{fig:upperLimitLifeTime} and 
Ref.~\cite{PastSearch}), it is difficult to compare precisely the 
limits with previous experiments due to the difference in collision 
centrality selection, acceptance, production model, rapidity search 
window and beam energy.  
Unfortunately, the theoretically expected yield in the forward region has  
yet to be calculated. 
 
In summary, the present search is the first one at RHIC energies in the  
forward region, and has produced no candidates.  The lifetime limits are  
less sensitive to experimental details than the limits in terms of  
production per event. Fig.~\ref{fig:upperLimitLifeTime} shows that this  
search puts much more stringent limits (around 0.1 ns, more than an order  
lower than before) on the lifetime of metastable strangelets. We emphasize  
that our search is generically sensitive to exotic objects with small  
charge-to-mass ratio in the kinematic regions shown in Fig.~\ref{fig:acceptance}. 
 
\begin{acknowledgments}         
We thank Sebastian White for consultations and help in building the STAR 
ZDC-SMDs. 
We thank the RHIC Operations Group and RCF at BNL, and the
NERSC Center at LBNL for their support. This work was supported
in part by the Offices of NP and HEP within the U.S. DOE Office 
of Science; the U.S. NSF; the BMBF of Germany; CNRS/IN2P3, RA, RPL, and
EMN of France; EPSRC of the United Kingdom; FAPESP of Brazil;
the Russian Ministry of Science and Technology; the Ministry of
Education and the NNSFC of China; IRP and GA of the Czech Republic,
FOM of the Netherlands, DAE, DST, and CSIR of the Government
of India; Swiss NSF; the Polish State Committee for Scientific 
Research; SRDA of Slovakia, and the Korea Sci. \& Eng. Foundation.
\end{acknowledgments}

\end{document}